# Tailoring topological Hall effect in SrRuO$_3$/SrTiO$_3$ superlattices

Seong Won Cho[a,b,1], Seung Gyo Jeong[c,1], Hee Young Kwon[d], Sehwan Song[e], Seungwu Han[b], Jung Hoon Han[c], Sungkyun Park[e], Woo Seok Choi[c,*], Suyoun Lee[a,f,*], Jun Woo Choi[d,*]

[a] Center for Neuromorphic Engineering, Korea Institute of Science and Technology, Seoul 02792, Korea

[b] Department of Materials Science and Engineering, Seoul National University, Seoul 08826, Korea

[c] Department of Physics, Sungkyunkwan University, Suwon 16419, Korea

[d] Center for Spintronics, Korea Institute of Science and Technology, Seoul 02792, Korea

[e] Department of Physics, Pusan National University, Busan 46241, Korea

[f] Division of Nano & Information Technology, Korea University of Science and Technology, Daejeon 34316, Korea

[1] These authors contributed equally to this work.

* Corresponding Authors. E-mail: choiws@skku.ac.kr (W. S. C.), slee_eels@kist.re.kr (S.L.), junwoo@kist.re.kr (J. W. C.)

**ABSTRACT**

Investigating the effects of the complex magnetic interactions on the formation of nontrivial magnetic phases enables a better understanding of magnetic materials. Moreover, an effective method to systematically control those interactions and phases could be extensively utilized in spintronic devices. $SrRuO_3$ heterostructures function as a suitable material system to investigate the complex magnetic interactions and the resultant formation of topological magnetic phases, as the heterostructuring approach provides an accessible controllability to modulate the magnetic interactions. In this study, we have observed that the Hall effect of $SrRuO_3/SrTiO_3$ superlattices varies nonmonotonically with the repetition number ($z$). Using Monte Carlo simulations, we identify a possible origin of this experimental observation: the interplay between the Dzyaloshinskii-Moriya interaction and dipole-dipole interaction, which have differing $z$-dependence, might result in a $z$-dependent modulation of topological magnetic phases. This approach provides not only a collective understanding of the magnetic interactions in artificial heterostructures but also a facile control over the skyrmion phases.

**1. Introduction**

Magnetic skyrmions are topologically nontrivial spin textures that induce intriguing spin dynamics and magneto-transport phenomena [1-8]. Recent experimental observations of skyrmions in numerous magnetic materials and thin film systems [1-6, 9-16] have restimulated the interest in two-dimensional (2D) spin textures, e.g. magnetic domains, which is a classic topic in magnetism. The ground state magnetic domain is determined by a competition of various magnetic interactions in the material system, including spin-spin exchange interaction, magnetic anisotropy energy, magnetic dipole-dipole interaction (DDI), and Dzyaloshinskii-Moriya interaction (DMI) [17-23]. In particular, the short-range DMI and long-range DDI favor the formation of small magnetic domains, such as stripes or skyrmions [1-6, 9-16, 19-23]. Distinct skyrmion phases, characterized by domain wall type and electrodynamics, are determined by the magnetic interaction (DMI or DDI) that predominantly stabilize the spin textures.

The systematic control of the two energy scales, i.e. the DMI and DDI, if possible, would provide a facile controllability over the spin-dependent properties and a fundamental understanding of the structure and stability of magnetic skyrmions. While some theoretical studies provided the energetics of topological spin textures considering the two interactions [22, 23], experimental studies involving the active control of these two energy terms are lacking. In this context, oxide superlattices (SLs) are ideal material systems allowing for deliberate design and modulation of the DMIs and DDIs simultaneously. Isostructural perovskite oxide SLs possess atomically well-defined interfaces owing to their strong covalent bonds between transition metals and oxygen. This offers a clear interpretation of the physical properties of the system and simultaneously provides the systematic tunability of material parameters via individual layer thickness, repetition number ($z$), stacking order, etc. In particular, heterostructures containing $SrRuO_3$ (SRO) serve as a platform to investigate the topological spin textures with an accessible lattice degree of freedom. SRO is an itinerant ferromagnet with appreciable spin-orbit coupling and exhibits various emergent functional properties including dimensionality-induced metal-insulator transition and adjustable topological phases [16, 24]. Interestingly, heterostructures with a few-nm-thick SRO frequently exhibit an unconventional Hall effect that is often interpreted as the topological Hall effect (THE) owing to the formation of skyrmions [16, 25-27]. The THE has been found to originate from a sizeable DMI, generated from interfacial inversion-symmetry breaking with finite spin-orbit coupling from Ru ions [28].

Herein, we report an emergent phenomenon resulting from the delicate competition between the DMI and DDI, in $SrRuO_3$/$SrTiO_3$ (SRO/STO) SLs. Fig. 1a shows the key strategy of the $z$-dependent magnetic interaction tuning in SRO/STO SLs via atomic-scale precision heterostructuring. We propose that the DMI and DDI can be systematically modulated by $z$ of the SLs. This results in the stabilization of two distinct skyrmion phases, leading to the $z$-dependent change of the magneto-transport properties of (SRO/STO)$_z$ SLs. We find that the magnitude of the skyrmion-phase-induced THE of (SRO/STO)$_z$ SLs shows a nonmonotonic dependence on $z$. The experimental observation is supported by Monte-Carlo simulations, which show that DMI-induced Néel-type skyrmions are stabilized for low-$z$, whereas DDI-induced Bloch-type skyrmions are stabilized for high-$z$. Thus, the atomic-scale precision heterostructuring of the controllable THE offers a viable method for modulating the topological spin structures.

## 2. Experimental methods

For the thin film growth, atomically controlled $[(SRO)_x|(STO)_y]_z$ ($[x|y]_z$) SLs with $x$ unit cell layers of SRO and $y$ unit cell layers of STO are repeated $z$ times using pulsed laser epitaxy on (001) STO substrates [24, 29-31]. Both SRO and STO layers are synthesized at 750 °C under 100 mTorr of oxygen partial pressure from a stoichiometric ceramic target, using a KrF laser (248 nm; IPEX-864, Lightmachinery). We use a laser fluence of 1.5 J·cm$^{-2}$ and a repetition rate of 5 Hz. For the stoichiometric film growth, we utilize high oxygen partial pressure and manipulate the number of atomic unit cells using a customized automatic laser pulse control system programmed in LabVIEW. Whereas oxygen vacancies can modify the magnetic properties of SRO [31-33], the high oxygen partial pressure of 100 mTorr used in the current study for the growth of the SLs leads to stoichiometric SRO layer with highly suppressed oxygen vacancies [24, 29].

We characterize the thickness of the SL period from high-resolution X-ray diffraction (XRD) $\theta$-$2\theta$ scans based on Bragg's law as follows:

$$\Lambda = \frac{\lambda}{2}(\sin\theta_n - \sin\theta_{n-1})^{-1},$$

where $\Lambda$, $n$, $\lambda$, and $\theta_n$ are the thickness of the SL period, SL order, X-ray wavelength, and $n$th-order peak position, respectively. All of the layers show a small thickness deviation below 1 unit cell thickness (~0.4 nm). The magnetic properties of the SLs are characterized by measuring the temperature-dependent magnetization ($M$ ($T$)) curves from 300 to 2 K under a 100 Oe out-of-plane magnetic field. The magnetic field-dependent magnetization ($M$ ($H$)) curves are obtained at several representative temperatures with an out-of-plane magnetic field.

The transport properties of the SLs, i.e., the longitudinal and Hall resistivities ($\rho_{xx}$ and $\rho_{xy}$, respectively), are measured based on the van der Pauw method, which gives a negligible contribution of $\rho_{xx}$ to $\rho_{xy}$. The resistance is measured by using a current source (K2612, Keithley) and a nano-voltmeter (K2182, Keithley) in a closed-cycle cryostat (CMag Vari.-9, Cryomagnetics) equipped with a 9T-superconducting magnet.

## 3. Results

*3.1. Structural and magnetic properties of the SLs*

We deliberately design SRO/STO SLs by systematically changing $z$. The atomic-scale precision control of the pulsed laser epitaxy enables the high-quality growth of the SRO/STO SLs (see Section 2) [24, 30, 34]. Fig. 1b-d shows the crystal structures of $[(SRO)_3|(STO)_8]_z$ ($[3|8]_z$, where $z = 2 \sim 50$) SLs with three (pseudocubic) unit cell layers of SRO and eight unit cell layers of STO alternatingly repeated $z$ times on (001) STO substrates. X-ray reflectivity (Fig. 1b) and $\theta$-$2\theta$ scans (Fig. 1c) consistently demonstrate the atomically defined $[3|8]_z$ SLs with different $z$ values, with the experimental thickness deviation of < 1 u.c. per supercell (see Supplementary Information (SI) Table S1). In particular, the SL peaks are clearly defined, indicating the coherent SL periods and conserved magnetic anisotropy of each sample. With increasing $z$, the SLs peaks sharpen systematically owing to the enhanced interference. Fig. 1d shows the XRD reciprocal space map (RSM) for the $[3|8]_{50}$ SL around the (103) Bragg reflection of the STO substrate. It represents the coherently strained state of the thickest SL to the substrate lattice. The $M(T)$ of the $[3|8]_{10}$ SL along the out-of-plane direction shows a typical ferromagnetic (FM) behavior with the transition temperature ($T_c$) of ~130 K (see SI Fig. S1). Note that the $T_c$ is smaller than that of bulk SRO (~150 K) owing to the thin layers within our SLs [24, 29, 30]. The out-of-plane $M(H)$ show square-like loops clearly supporting FM ordering below $T_c$ (see SI Section S2 for $M(H)$ and $M(T)$ data).

*3.2. Transport properties*

Fig. 1e shows the temperature-dependent resistivity $\rho_{xx}(T)$ of the $[3|8]_z$ SLs. The overall qualitative characteristic $\rho_{xx}(T)$ of the SLs closely resemble those of the SRO, including a kink at $T_c = $ ~130 K. The only difference is an upturn below ~30 K, possibly originating from the dimensionality crossover previously observed for $[2|8]_{10}$ SL or disorder-induced localization reported in ultrathin SRO layers [35]. Fig. 1f shows the magnetoresistance, MR = [$\rho_{xx}$ (0) − $\rho_{xx}$ (4.5 T)] / $\rho_{xx}$ (0) × 100 (%), as a function of *T*. All SLs show a double-peak feature with one peak at ~135 K and the other at ~75 K. A similar double-peak structure in the MR vs. *T* curve was reported in SrRuO$_3$-SrIrO$_3$ bilayer systems [25], although the origin was not determined. While the peak at ~135 K can be attributed to FM ordering around $T_c$, the origin of the peak at ~75 K is unclear at the moment. Nevertheless, in the SrRuO$_3$-SrIrO$_3$ bilayer system, this temperature was defined as the temperature of the maximum THE [25]. We observe a similar enhancement in the THE around this temperature in our SLs, as will be shown in Section 3.3.

*3.3. Nonmonotonic z-dependence of THE in [3|8]$_z$ SLs*

A typical Hall resistivity ($\rho_{xy}$) of the [3|8]$_{10}$ SL as a function of the external magnetic field (*H*) measured at 60 K is shown in the top panel of Fig. 2a. Similar to material systems that feature skyrmions [16, 25, 26, 36-38], a hump structure is observed around the coercive field ($H_c$). Note that some studies attribute the hump signal to inhomogeneity within the film, i.e., multiple regions with opposite Hall effect signs [39]. However, the absence of any SRO thickness- or temperature-dependent change in the sign of the AHE in our measurements (see SI Figs. S4 and S5), along with the symmetric 'minor-loops' (see SI Fig. S7), confirm that the hump in our SL samples most probably originates from the existence of skyrmions [38], eliminating the possibility of film inhomogeneity as the origin of the hump discussed in some previous literatures [39]. See SI Section S6 for further discussion regarding evidence for the THE-origin of the hump signals in our SLs. Furthermore, the THE is absent in the SLs with thicker SRO layers (see SI Fig. S4), revealing the interfacial nature of the DMI in our SLs.

To quantitatively analyze the *z*-dependence of the [3|8]$_z$ SLs, we isolate the contribution of the THE from the measured $\rho_{xy}$ using Eq. (1) as follows:

$$\rho_{\text{tot}} = \rho_{\text{OHE}} + \rho_{\text{AHE}} + \rho_{\text{THE}} = R_0 H + R_S M + R_0 B_{\text{eff}}. \tag{1}$$

Here, $\rho_{OHE}$, $\rho_{AHE}$, and $\rho_{THE}$ represent the ordinary, anomalous, and topological Hall resistivities, respectively. In addition, $R_0$, $R_S$, and $B_{eff}$ denote the ordinary Hall constant, anomalous Hall constant, and effective magnetic field, respectively [25]. It is noteworthy that $R_0$ is not constant but depends on the $H$-field (see SI Fig. S8). Such a nonlinear dependence of $\rho_{OHE}$ on the $H$-field might imply the coexistence of multiple types of carriers. Therefore, the contribution of the ordinary Hall effect is subtracted by fitting the data outside the coercive field to the two-band model of $\rho_{xy}$ [40]. Additionally, $\rho_{THE}$ of the SL samples is obtained by subtracting $\rho_{AHE}$ that is assumed as a step function (see SI Section S7 for details on obtaining $\rho_{THE}$ from $\rho_{xy}$) [41].

Upon observing $\rho_{xy}$, one can immediately notice an unexpected nonmonotonic $z$-dependence of the THE contribution. Such unique characteristics of our SLs are also evident in the raw data of $\rho_{xy}$ ($H$) curves which show that the hump feature is most prominent for the $z$ = 2 and 10 SLs (see SI Fig. S5). We note that the large hump features are consistently reproducible in multiple samples with the same SL configuration (see SI Fig. S6, e.g., for $z$ = 10 SLs). The obtained $\rho_{THE}$ is plotted as a function of $H$ at several selected temperatures for different $z$ values (Fig. 2b). Two intriguing features are discovered in the color-map plots in the ($H$, $T$) space (Fig. 2b): (1) the nonmonotonic dependence of $\rho_{THE}$ on $z$ and (2) the independence of $T_{max}$ (the temperature of the maximum THE) on $z$, where $\rho_{THE}$ is maximum. To capture the $z$-dependence more effectively, $\rho_{THE}$ at 60 K (~$T_{max}$) is plotted as a function of $z$ (Fig. 2c). $\rho_{THE}$ shows a clear decreasing trend with $z$ increasing from two to five, above which $\rho_{THE}$ recovers its strength at $z$ = 10 and 20. Note that the nonmonotonic $z$-dependence of THE, plotted for the $T$ = 60 K data in Fig. 2c, is evident at all $T$ (see Fig. 2b). Fig. 2d shows $\rho_{THE}$ as a function of $T$, verifying the aforementioned independence of $T_{max}$ on $z$. This suggests that the skyrmion formation energy is comparable irrespective of $z$. It is noteworthy that $T_{max}$ is approximately 60 K, slightly lower than the temperature corresponding to the MR peak (~75 K) as mentioned in Fig. 1f.

*3.4. Micromagnetic simulations*

To explain the nonmonotonic *z*-dependence of $\rho_{THE}$ shown in section 3.3, we perform a simulation study on the spin configurations taking into account the *z*-dependence of the DMI and DDI, which are the two principal interactions favoring magnetic domain formation. Note that the former is a short-range interaction while the latter is a long-range interaction. Micromagnetic simulations are performed using the Monte-Carlo method with a 2D square grid system and the Heisenberg model, which has been used in previous studies to investigate 2D magnetic domains [20]. We consider an SL system where the DMI and long-range DDI scale with $1/z$ and $z$, respectively. These are natural assumptions for film systems with interfacial DMI (see SI Section S8 for elaboration on the *z*-dependent energy scaling and origin of the interfacial DMI in an SL system, along with details of the simulation methods). Also note that in earlier experimental studies on magnetic multilayer systems (repetitions of ferromagnetic/nonmagnetic layers) [12, 13, 42, 43], all the perpendicularly magnetized ferromagnetic layers within the system were strongly ferromagnetically interlayer coupled across several-nm-thick non-magnetic layers, which resulted in the formation of 2D spin textures such as skyrmions. Therefore, in our model, we assume a strong perpendicular anisotropy and strong interlayer coupling, such that all the ferromagnetic layers in the SL behave collectively as a single magnetic entity, i.e., the domains in our SL system are effectively 2D spin textures (see SI Section S8 for more details).

Fig. 3a-c show the calculated spin configurations as the strengths of the DMI and DDI are varied. Starting with a small *z* (large DMI/small DDI), *z* is gradually increased such that the DMI and DDI decrease and increase, respectively (Fig. 3a). Multiple bubble-shaped magnetic domains (i.e. skyrmions) exist for significantly large DMIs for a small *z* ($z = 2$, case (i) in Fig. 3c). The in-plane *M* within all the skyrmion domain walls are identical and pointing into the skyrmion core; this implies that these domains are homochiral Néel-type skyrmions. As *z* is increased ($z = 5$), the skyrmions disappear owing to the decrease in the DMI and the increase in the DDI (Fig. 3c-(ii)). For these intermediate values of the DMI and DDI, neither interaction is sufficiently strong for stabilizing the skyrmions. That is, neither domain formation preferring energy term is large enough to overcome the combined effect of the exchange interaction and magnetic-field-induced Zeeman energy, both of which prefer a parallel spin alignment, such that single domains form at these intermediate $z$ values ($z = 4 - 8$, see SI Fig. S9). Meanwhile, a multi-skyrmion state reappears when $z$ is further increased ($z = 10$, Fig. 3c-(iii)), owing to the increased DDI. We discover that (1) the in-plane $M$ within the skyrmion domain walls are winding around the skyrmion core and (2) skyrmions with clockwise and counter-clockwise winding directions coexist. This shows that the DDI-stabilized skyrmions in Fig. 3c-(iii) are Bloch-type with mixed chirality, which is expected given that the DDI is a nonchiral interaction that prefers Bloch-type domain walls. As $z$ is further increased ($z = 20$), the skyrmion phase once again disappears as the magnetic stripe phase form (Fig. 3c-(iv)) owing to the large DDI, which prefers the stripe phase with 50:50 (up:down) magnetizations.

Evaluation of the number of skyrmions ($N_{sk}$) in the simulated domain images shows a nonmonotonic $z$-dependence with peaks at $z = 2$ and 10 (Fig. 3b). As discussed previously, the THE is a key feature of the skyrmion phase; the THE signal is proportional to the skyrmion density [16, 25, 43]. Therefore, the qualitative agreement between the simulated number of skyrmions (Fig. 3b) and $\rho_{THE}$ (Fig. 2c) implies that the nonmonotonic $z$-dependence of $\rho_{THE}$ is possibly due to the $z$-dependent modulation of the existence and stability of the skyrmion phases. In the low-$z$ region, a DMI-stabilized Néel-type multi-skyrmion phase is formed, resulting in a large $\rho_{THE}$. Conversely, in the high-$z$ region, the DDI-stabilized Bloch-type multi-skyrmion phase leads to a large $\rho_{THE}$. Note that modifying the chirality alone would not change the sign of $\rho_{THE}$ if the polarity is not switched [16, 43]. Therefore, the Bloch-type skyrmions with mixed chirality and fixed polarity (Fig. 3c-(iii)) would result in a large $\rho_{THE}$, similar to the homochiral

Néel-type skyrmions (Fig. 3c-(i)). From the simulations, we suspect that the disappearance of the THE at intermediate and very high $z$ values is due to the absence of skyrmions and the formation of stripe domains, respectively. While these simulation results provide one possible scenario that can explain the nonmonotonic $z$-dependence of the THE, further measurements such as high-resolution magnetic domain imaging are required to unambiguously confirm the exact origin of the experimental results.

## 4. Discussion

Note that the two topological phases can exist in an identical material system only by tuning $z$ of the SLs. Whereas the simulation results are applicable to single-layer films in principle, typically the magnetic anisotropy of single-layer films is not an independent parameter as it can vary significantly with the thickness ($t$). In contrast, the magnetic anisotropy does not change with $z$ in our SL system. Specifically, the coercive field being invariant over a wide range of $z$ = 2 – 50 (see SI Fig. S5) demonstrates that the magnetic anisotropy is largely unchanged with $z$. This confirms that SLs with constant unit layer $t$ and varying $z$ consist the most appropriate material system to elucidate the competition between the DMI and DDI selectively while maintaining the other energy terms in the Hamiltonian, such as the exchange interaction and magnetic anisotropy constant.

## 5. Conclusion

In summary, we have observed that the THE of [SRO/STO]$_z$ SLs vary nonmonotonically with $z$. One possible scheme that might explain this experimental observation is the interplay between two magnetic interactions, the short-range DMI and the long-range DDI, leading to distinct topological magnetic phases depending on $z$. The SRO/STO SL exemplifies a material system that demonstrates the competition between the DMI and DDI, and its effect on the topological spin textures. Our atomic-scale strategy demonstrates the facile controllability of the THE for a better understanding of the topological spin textures and the possible realization of next-generation spin topology based devices.

**Acknowledgements**

This work was supported by the Basic Science Research Programs through the National Research Foundation of Korea (NRF) funded by the Ministry of Science and ICT (MSIT) (NRF-2021R1A2C2011340, NRF-2018R1D1A1B07045663, NRF-2019M3F3A1A02072175, and NRF-2020R1A5A1104591), and the KIST Institutional Program (2E30141, 2E31032).

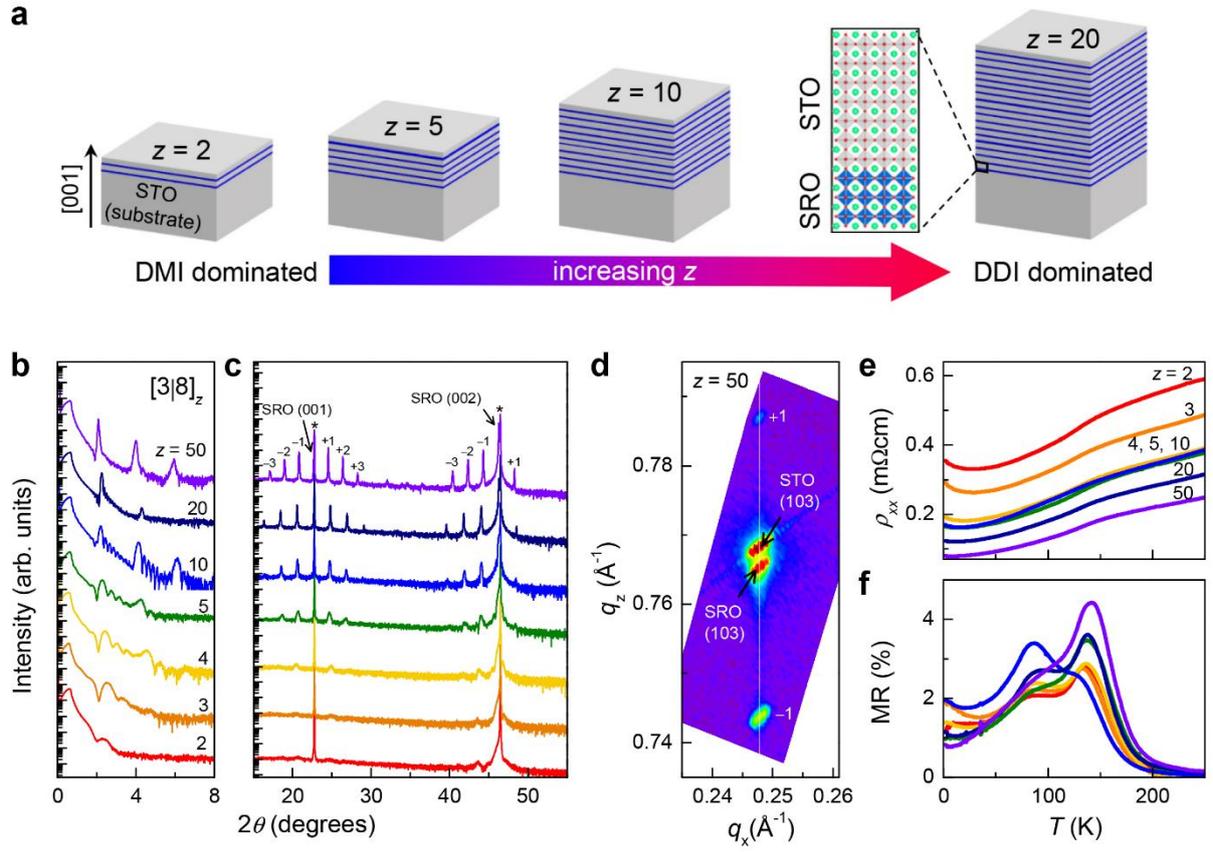

**Fig. 1.** Scheme for tailoring topological magnetic phases in SLs, and basic material characteristics of atomically controlled $[3|8]_z$ SLs. (a) Schematic representation of $z$-dependent DMI and DDI of atomically-designed SRO/STO SLs. With increasing $z$, the dominant magnetic interaction changes from DMI to DDI. (b) X-ray reflectivity and (c) XRD $\theta$-$2\theta$ scan shown for $[3|8]_z$ SLs grown on STO substrates. Asterisks (*) indicate Bragg peak positions of (00$l$) STO substrate. Identical $2\theta$ angles of SL peaks ($\pm n$) for different samples indicate coherent supercell structures composed of three-unit-cell layers of SRO and eight-unit-cell layers of STO. With increasing $z$, the intensity of SLs peaks enhances gradually. (d) RSM of $[3|8]_{50}$ SL, around (103) Bragg reflection of STO substrate, indicating a fully strained state. $T$-dependent (e) $\rho_{xx}$ and (f) MR have the same color code for the same $z$. $\rho_{xx}(T)$ is determined by the product of the measured sheet resistance and the total thickness of the SRO layer within the SLs.

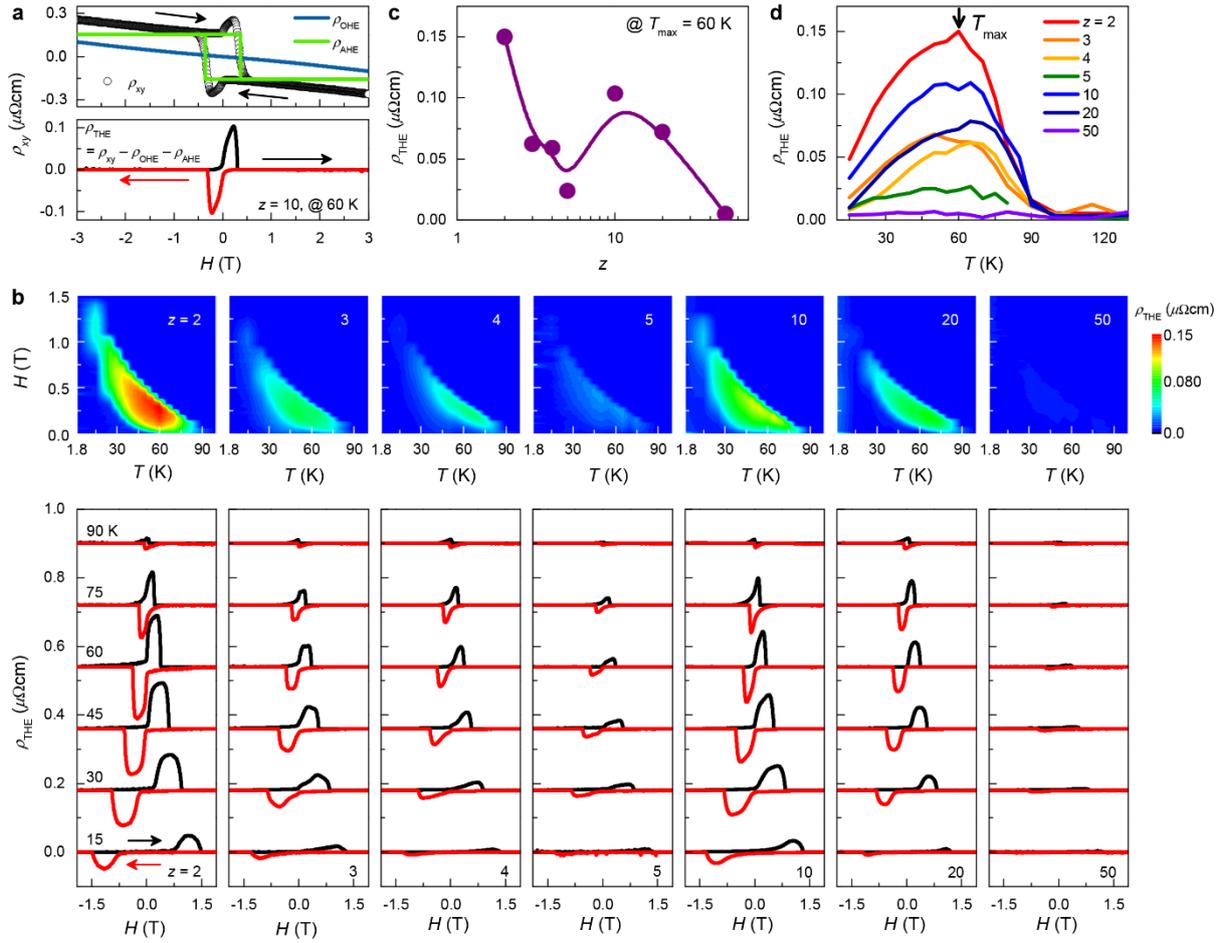

**Fig. 2.** Nonmonotonic *z*-dependence of topological Hall effect in [3|8]$_z$ SLs via atomic-scale precision heterostructuring. (a) (top) Hall resistivity ($\rho_{xy}$), (bottom) extracted THE signal after subtracting $\rho_{OHE}$ and $\rho_{AHE}$. (b) (top) Contour plots of $\rho_{THE}$ as a function of *H*-field and *T*, (bottom) $\rho_{THE}$ (*H*) curves at several selected *T*. (c) *z*-dependence of $\rho_{THE}$ at 60 K. (d) *T* dependence of $\rho_{THE}$ with varying *z*. The arrows indicate *H*-field directions.

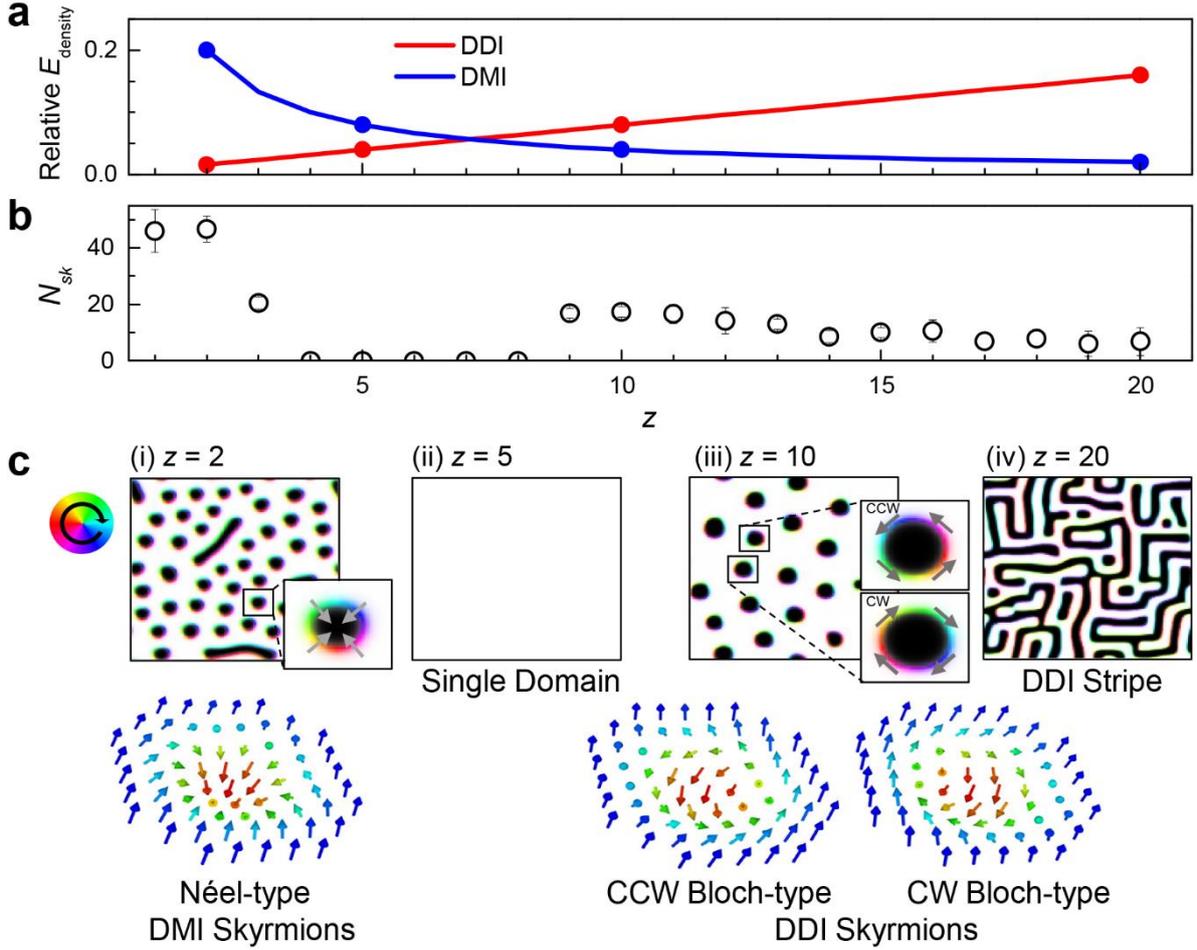

**Fig. 3.** Monte-Carlo simulations of spin texture with DMI and DDI modulations. (a) $z$-dependent DMI and DDI used for Monte-Carlo simulations. (b) $z$-dependence of the number of skyrmions ($N_{sk}$). Each $N_{sk}$ is averaged from 10 simulated images. (c) Spin configurations under a constant out-of-plane magnetic field and for different $z$ values. Schematic diagrams of Néel-, counter-clockwise (CCW) Bloch, and clockwise (CW) Bloch-type skyrmions are shown below. The black/white contrasts correspond to out-of-plane spin directions, and the color scale shows the in-plane spin directions. For the skyrmion schematics, the color scale corresponds to the out-of-plane spin directions for a better visualization of the skyrmion core. Full $z$-dependent domain evolution is shown in SI Fig. S9.